\documentclass[12pt]{article}
\usepackage{fullpage}
\usepackage[dvips]{epsfig}

\def\##1{{\bf #1}}
\def\=#1{\underline{\underline{#1}}}

\def\+#1{\underline{\bf #1}}
\def\*#1{\underline{\underline{\bf #1}}}

\def\r#1{(\ref{#1})}
\def\l#1{\label{#1}}
\def\c#1{\cite{#1}}

\def\le{\left(}
\def\ri{\right)}
\def\les{\left[}
\def\ris{\right]}
\def\lec{\left\{}
\def\ric{\right\}}

\def\.{\mbox{ \tiny{$^\bullet$} }}

\def\muo{\mu_{\scriptscriptstyle 0}}

\def\ko{k_{\scriptscriptstyle 0}}

\def\eps{\epsilon}

\begin{document}

\Large
\begin{center}
{\bf Correlation Length Facilitates \\ Voigt Wave Propagation}

\vspace{10mm} \large

Tom G. Mackay\footnote{Permanent address: School of Mathematics,
University of Edinburgh, Edinburgh EH9 3JZ, UK.\\ Fax: + 44 131
650 6553; e--mail: T.Mackay@ed.ac.uk.}
 and Akhlesh  Lakhtakia

\end{center}

\vspace{4mm}

\normalsize

 \noindent CATMAS~---~Computational \& Theoretical
Materials Sciences Group\\ Department of Engineering Science and
Mechanics, Pennsylvania State University\\ University Park, PA
16802--6812, USA

\vspace{5mm}

\begin{abstract}

Under certain circumstances, Voigt waves can propagate in a
biaxial composite medium even though the  component material
phases individually do not support Voigt wave propagation. This
phenomenon is considered within the context of the
strong--permittivity--fluctuation theory. A generalized
implementation of the theory is developed in order to explore the
propagation of Voigt waves in any direction. It is shown that the
correlation length~---~a parameter characterizing the
distributional statistics of the component material
phases~---~plays a crucial role in facilitating the propagation of
Voigt waves in the homogenized composite medium.

\end{abstract}


\section{Introduction}

A defining characteristic of metamaterials is that they exhibit
behaviour which is not exhibited by their component phases
\c{Wals}. A prime example is provided by homogenized composite
mediums (HCMs) which support Voigt wave propagation despite their
component material phases  not doing so \c{ML03}. Although they were
discovered over 100 years ago \c{Voigt},   Voigt waves are not
widely known in the optics/electromagnetics community. However,
they have recently become the subject of renewed interest
\c{Lakh98,Berry}, and the more so in light of advances in complex composite
mediums \c{Singh}.

A Voigt wave is an anomalous plane wave which
can develop in certain anisotropic mediums when the associated
propagation matrix is not diagonalizable \c{GL01}. The
unusual property of a Voigt wave is that its amplitude is
linearly dependent upon the propagation distance \c{Khap}.

In a recent study, the  Maxwell Garnett and Bruggeman
homogenization formalisms were applied to show that Voigt waves
may propagate in biaxial HCMs provided that their component
material phases are inherently dissipative \c{ML03}. However, the
Maxwell Garnett and Bruggeman formalisms~---~like many
widely--used homogenization formalisms \c{L96}~---~do not take
account of coherent scattering losses. The
strong--permittivity--fluctuation theory (SPFT) provides an
alternative approach in which  a more comprehensive description of
the distributional statistics of the component material phases is
accommodated \c{TK81, Genchev}. In the bilocally--approximated
implementation of the SPFT, a two--point covariance function and
its associated correlation length $L$ characterize the component
phase distributions. Coherent interactions between pairs of
scattering centres within a region of linear dimensions $L$ are
thereby considered in the SPFT, but scattering centres separated
by distances much greater than $L$ are assumed to act
independently. Thus, the SPFT provides an estimation of coherent
scattering losses, unlike the Maxwell Garnett and Bruggeman
formalisms. In fact, the bilocally--approximated SPFT gives rise
to the Bruggeman homogenization formalism  in the limit $L
\rightarrow 0$ \c{MLW00}.

In the following sections, we consider Voigt wave propagation in a
biaxial two--phase HCM within the context of the SPFT. A
generalized SPFT implementation is developed in order to explore
the propagation of Voigt waves in any direction. The unit
Cartesian vectors are denoted as $\hat{\#x}$, $\hat{\#y}$ and
$\hat{\#z}$. Double underlined quantities are 3$\times$3 dyadics.
The wavenumber of free space (i.e., vacuum) is $\ko$.

\section{Homogenization background}

\subsection{Component phases}

The propagation of Voigt waves in a two--phase homogenized
composite medium (HCM) is  investigated. Both component material phases are
taken as uniaxial dielectric mediums. Therefore, they do not individually
support Voigt wave propagation \c{GL01}.

Let the components
phases~---~labelled as $a$ and $b$~---~be characterized by the
relative permittivity dyadics
\begin{equation}
\left.
\begin{array}{l}
\=\eps_{a} =
\=R_{\,z} (\varphi) \. \les \eps^x_a \, \hat{\#x} \, \hat{\#x} +
\eps_a \,
\le  \, \hat{\#y} \, \hat{\#y} + \hat{\#z} \, \hat{\#z} \, \ri  \ris
\. \=R^T_{\,z} (\varphi)
\\
\vspace{-3mm}
\\
\=\eps_b =   \eps^x_b \, \hat{\#x} \, \hat{\#x} +
\eps_b \,
\le  \, \hat{\#y} \, \hat{\#y} + \hat{\#z} \, \hat{\#z} \, \ri
\end{array}
\right\},
\l{permab}
\end{equation}
respectively. The preferred axis of component material phase $a$ is rotated
under the action of
\begin{equation}
\=R_{\,z} (\varphi) =
 \cos \varphi
\le  \, \hat{\#x} \, \hat{\#x} + \hat{\#y} \, \hat{\#y} \, \ri +
 \sin \varphi
\le  \, \hat{\#x} \, \hat{\#y} - \hat{\#y} \, \hat{\#x} \, \ri
+  \hat{\#z} \, \hat{\#z} \,,
\end{equation}
to lie in the $xy$ plane at an angle $\varphi$ to the $x$ axis,
whereas the preferred axis of component phase $b$ is aligned with
the $x$ axis, without loss of generality. The superscript $^T$ indicates the
transpose operation.

Let the regions occupied by component phases $a$ and $b$ be
denoted by $V_a$ and $V_b$, respectively. The component phases are
 randomly distributed such that all space $V = V_a \cup V_b$.
Spherical microstructural geometries, with characteristic length
scales which are small in comparison with electromagnetic
wavelengths, are assumed for both component phases. The
distributional statistics of the component phases are described in
terms of moments  of the characteristic functions
\begin{equation}
\Phi_{ \ell}(\#r) = \left\{ \begin{array}{ll} 1\,, & \qquad \#r
\in
V_{\, \ell}\,,\\ & \qquad \qquad \qquad \qquad \qquad \qquad (\ell=a,b) . \\
 0\,, & \qquad \#r \not\in V_{\, \ell}\,, \end{array} \right.
\end{equation}
The volume fraction $f_\ell$ of phase $\ell$  is given by
the first statistical moment of
 $\Phi_{\ell}$;
 i.e., $\langle \, \Phi_{\ell}(\#r) \, \rangle = f_\ell$ .
 Clearly,
 $f_a + f_b = 1$. The second statistical moment of $\Phi_{\ell}$
 provides
a two--point covariance function. We adopt  the physically motivated form
\c{TKN82}
\begin{equation}
\langle \, \Phi_\ell (\#r) \, \Phi_\ell (\#r')\,\rangle = 
\left\{
\begin{array}{lll}
\langle \, \Phi_\ell (\#r) \, \rangle \langle \Phi_\ell
(\#r')\,\rangle\,, & & |  \#r - \#r' | > L \\ && \hspace{25mm}, \\
\langle \, \Phi_\ell (\#r) \, \rangle \,, && |  \#r - \#r' | \leq L
\end{array}
\right.
 \l{cov}
\end{equation}
wherein 
$L>0$ is the correlation length. 
Thus, coherent interactions between a  scattering centre
located at $\#r$ and another located at $\#r'$ are accommodated provided that
$ |  \#r - \#r' | \leq L$. However, if $ |  \#r - \#r' | > L $ then
the scattering centre at  $\#r$ is presumed to act independently of the
scattering centre at $\#r'$. 
In implementations of the SPFT,
the precise form of the covariance function is relatively unimportant
to estimate the  constitutive parameters of the HCM
\c{MLW01b}.

\subsection{Homogenized composite medium}

Since the preferred axes of the uniaxial component phases are
not generally aligned,  the HCM is a biaxial dielectric medium.
We confine ourselves to the bilocally approximated
 SPFT in order to estimate the HCM relative permittivity dyadic
\begin{eqnarray}
\=\eps_{\,ba}  &=& \eps^x_{ba} \, \hat{\#x} \,  \hat{\#x} +
\eps^y_{ba} \, \hat{\#y} \,  \hat{\#y} + \eps^z_{ba} \, \hat{\#z}
\,  \hat{\#z} + \eps^t_{ba} \, \le \hat{\#x} \,  \hat{\#y} +
\hat{\#y} \,  \hat{\#x} \ri.
 \l{ba2}
\end{eqnarray}
The SPFT is based upon  iterative refinements of a comparison
medium. The relative permittivity dyadic of the comparison medium,
namely,
\begin{eqnarray}
\=\eps_{\,Br}  &=& \eps^x_{Br} \, \hat{\#x} \,  \hat{\#x} +
\eps^y_{Br} \, \hat{\#y} \,  \hat{\#y} + \eps^z_{Br} \, \hat{\#z}
\,  \hat{\#z} + \eps^t_{Br} \, \le \hat{\#x} \,  \hat{\#y} +
\hat{\#y} \,  \hat{\#x} \ri, \l{Bruggeman}
\end{eqnarray}
is provided by
 the Bruggeman homogenization formalism \c{MLW00}.

\subsection{Depolarization and polarizability dyadics}

The depolarization dyadic $\=D$ is central to  both the  Bruggeman formalism and the
SPFT. It provides the electromagnetic response of
a infinitesimally small spherical exclusion region, immersed in a
homogeneous background. For the comparison medium with relative
permittivity dyadic $\=\eps_{\,Br}$, the corresponding
depolarization dyadic is given by\c{M97}
\begin{eqnarray}
\=D &=& \frac{1}{ i\, \omega \, 4 \pi} \, \int^{2\pi}_0 \; d\phi
\, \int^\pi_0\;  d\theta\;
\frac{\sin \theta }{\Gamma^\phi_{Br} \, \sin^2 \theta + \eps^z_{Br} \cos^2 \theta }
\, \hat{\#q}\, \hat{\#q} \,, \l{depol}
\end{eqnarray}
 wherein
\begin{equation}
\Gamma^\phi_{Br} = \eps^x_{Br} \, \cos^2 \phi + \eps^y_{Br} \, \sin^2 \phi + 2 \eps^t_{Br}
\, \sin \phi \cos \phi,
\end{equation}
and $\hat{\#q} = \sin \theta \cos \phi \,  \hat{\#x}+ \sin \theta \sin \phi\,  \hat{\#y}+
\cos \theta \,  \hat{\#z}$ is the unit position vector.

A related  construction, much used in homogenization formalisms,
 is the polarizability density dyadic $\=\chi_{\,\ell}$. It is defined here
  as
\begin{equation}
\=\chi_{\,\ell} = -i\,\omega\,\le\,\=\eps_{\,\ell} -
\=\eps_{\,Br}\,\ri\. \les\, \=I + i \omega \, \=D\.\le\,
\=\eps_{\,\ell} - \=\eps_{\,Br}\,\ri\,\ris^{-1}, \hspace{20mm} (\ell
=a,b). \l{X_def}
 \end{equation}

\subsection{The bilocally approximated SPFT}

After accommodating higher--order distributional statistics, the
bilocally approximated SPFT estimate
\begin{equation}
\=\eps_{\,ba} = \=\eps_{\,Br} - \frac{1}{i \,\omega}\,\le\,\=I +
\=\Sigma_{\,ba} \. \=D \,\ri^{-1}\.\=\Sigma_{\,ba}\,
\end{equation}
is derived  \c{MLW00}. The \emph{mass operator} \c{Frisch} term
\begin{equation}
\=\Sigma_{\,ba} = \le\,\=\chi_{\,a} -
\=\chi_{\,b}\,\ri\.\=P\.\le\,\=\chi_{\,a} - \=\chi_{\,b}\,\ri
\end{equation}
is specified in terms of the  principal value integral
\begin{equation}
\=P = \mathcal{P} \,\int_{\sigma \leq L} \; d^3 \#R \: \;
\=G_{\,Br}(\#R ), \l{W_def}
\end{equation}
with
$\#R =  \#r - \#r'$ and 
  $\=G_{\,Br} (\#R)$ being
 the unbounded dyadic Green
function of the comparison medium. A surface integral
representation of $\=P$ is established in the Appendix.
Thereby, we see that $\=\eps_{\,ba}$ has a complex dependency
upon the correlation length $L$, with $\=\eps_{\,ba}$ 
becoming  equal to 
$\=\eps_{\,Br}$ in the limit $L \rightarrow 0$.

\section{Voigt wave propagation}

In order to explore the possibility of Voigt wave
propagation~---~in any direction~---~it is expedient to consider
the HCM relative permittivity dyadic \r{ba2} in the rotated
coordinate frame specified by the three Euler angles $\alpha$,
$\beta$ and $\gamma$  \c{Arfken}. On rotation, $\=\eps_{\,ba}$
transforms to
\begin{eqnarray}
\={\hat{\eps}}_{\,ba} (\alpha, \beta, \gamma ) &=&
 \=R_{\,z}(\gamma)\.\=R_{\,y}(\beta)\.\=R_{\,z}(\alpha)\.\=\eps_{\,ba}\.
 \=R^T_{\,z}(\alpha)\.\=R^T_{\,y}(\beta)\.\=R^T_{\,z}(\gamma) \l{HCMrot2} \\
 &=& \eps_{11} \,
\hat{\#x} \, \hat{\#x} + \eps_{22} \, \hat{\#y} \,  \hat{\#y} +
\eps_{33} \, \hat{\#z} \,  \hat{\#z} +  \eps_{12} \, \le \hat{\#x}
\, \hat{\#y} + \hat{\#y} \,  \hat{\#x} \ri \nonumber \\ &&  +
\eps_{13} \, \le \hat{\#x} \, \hat{\#z} + \hat{\#z} \,  \hat{\#x}
\ri + \eps_{23} \, \le \hat{\#y} \, \hat{\#z} + \hat{\#z} \,
\hat{\#y} \ri,
 \l{HCMrot1}
\end{eqnarray}
wherein
\begin{equation}
\=R_{\,y} (\beta)= \cos \beta \le  \, \hat{\#x} \, \hat{\#x} +
\hat{\#z} \, \hat{\#z} \, \ri +
 \sin \beta
\le  \, \hat{\#z} \, \hat{\#x} - \hat{\#x} \, \hat{\#z} \, \ri +
\hat{\#y} \, \hat{\#y} \,.
\end{equation}

Two conditions must be satisfied in order for Voigt waves to
propagate along the $z$ axis \c{GL01}:
\begin{itemize}  \item[(i)] $Y (\alpha, \beta, \gamma ) = 0$, and
\item[(ii)] $W (\alpha, \beta, \gamma ) \neq 0$.
\end{itemize}
  In terms of the relative permittivity scalars on the right side of \r{HCMrot1}
   (which are all implicit functions of $\alpha$, $\beta$ and $\gamma$),
\begin{eqnarray}
Y(\alpha, \beta, \gamma ) &=& \eps^4_{13} + \eps^4_{23} -2
\eps_{23}\eps_{33}
 \les \, 2 \eps_{12}
\eps_{13} - \le \, \eps_{11} - \eps_{22}\, \ri \eps_{23}\,\ris +
\les \le \, \eps_{11}-\eps_{22}\,\ri^2 + 4 \eps^2_{12}\,\ris \,
\eps^2_{33} \nonumber \\ && + 2 \eps_{13} \lec \, \eps^2_{23}
\eps_{13} - \les \, 2 \eps_{12}\eps_{23} + \le \, \eps_{11} -
\eps_{22} \, \ri \, \eps_{13}\,\ris \eps_{33}\,\ric
\end{eqnarray}
and
\begin{equation}
W(\alpha, \beta, \gamma ) = \eps_{12} \eps_{33} - \eps_{13}
\eps_{23}\,.
\end{equation}

\section{Numerical results}

The numerical calculations proceed in two stages: Firstly,
$\=\eps_{\,ba}$ is estimated using the bilocally approximated SPFT
for a representative example; secondly,
 the quantities
$Y(\alpha, \beta, \gamma )$ and $W(\alpha, \beta, \gamma )$ are calculated as functions of the Euler angles.
In particular, the $(\alpha,
\beta, \gamma )$ angular coordinates of the zeros of $Y$, and the
corresponding values of $W$ at  those $(\alpha, \beta, \gamma )$
coordinates, are sought. The $\gamma$ angular coordinate
is disregarded since propagation parallel to the $z$ axis (of
the rotated coordinate system) is independent of rotation about
that axis.

The following constitutive parameters were selected
for the component phases $a$ and $b$ for all results
presented here:
 \begin{equation}
 \left.
 \begin{array}{ll}
  \eps^a_x =  2.5 +  i \, 0.1 \,\delta   , &
\eps^a =
 6.8 +   i\, 0.25 \,\delta \\ \vspace{-2.5mm} & \\
\eps^b_x = 3.6 +  i\, 2.5 \,\delta, & \eps^b =   2.2 + i\, 1.4
\,\delta
\end{array}
\right\},
\end{equation}
with the dissipation parameter  $\delta \in [0,1]$.
 The  volume  fraction $f_a = 0.3 $ for all calculations.

\subsection{HCM constitutive parameters}

\begin{figure}[!ht]
\centering \psfull \epsfig{file=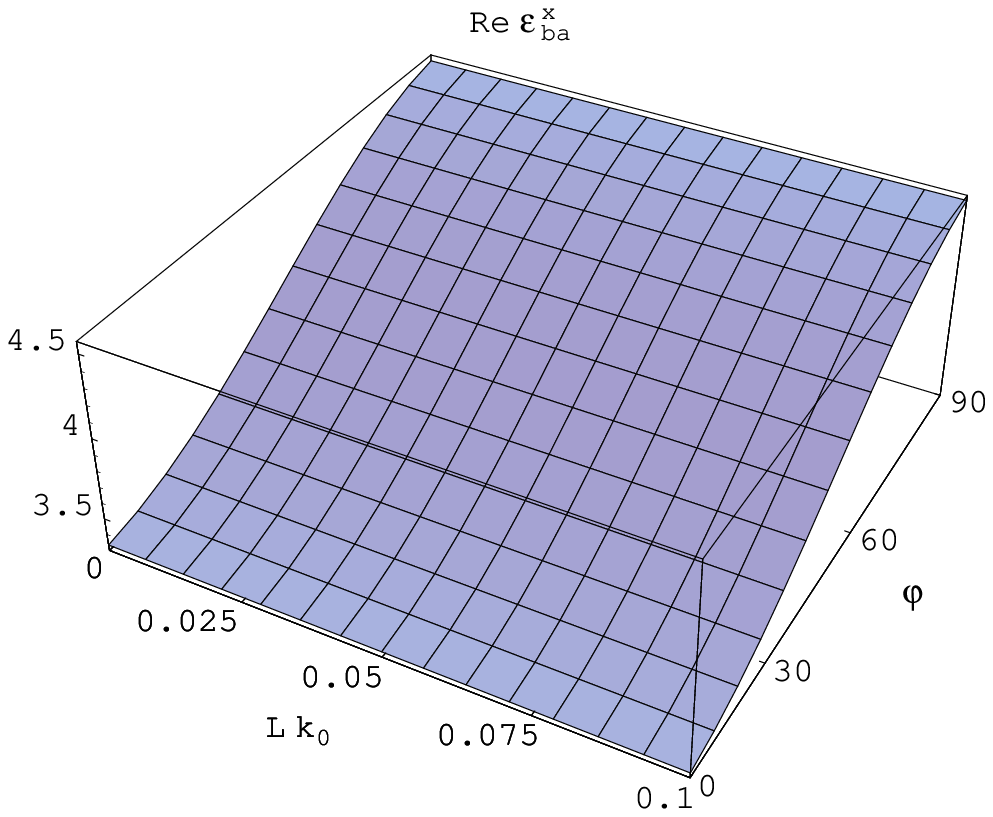,width=2.5in} \hfill
\epsfig{file=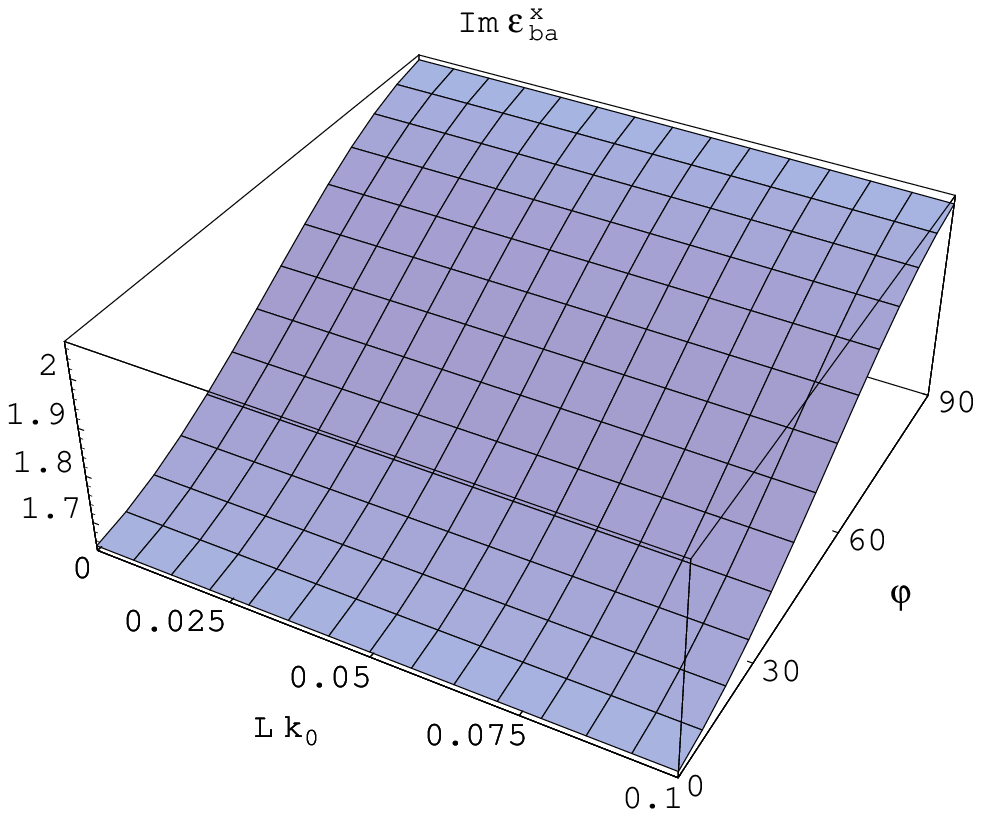,width=2.5in}
\epsfig{file=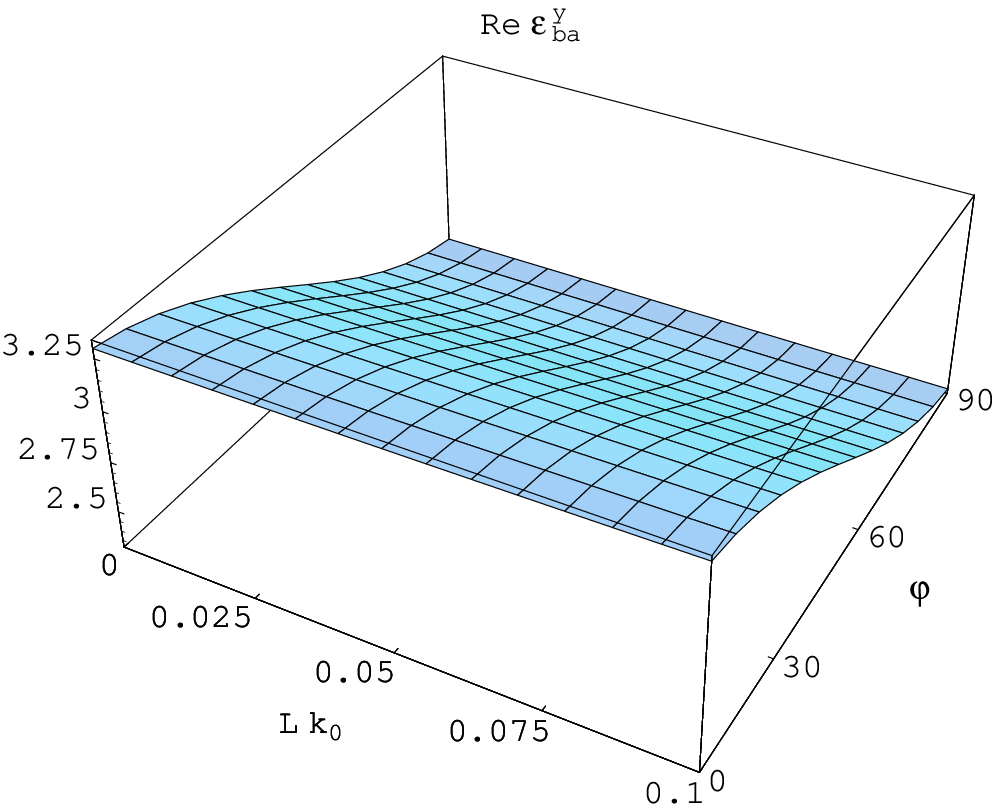,width=2.5in} \hfill
\epsfig{file=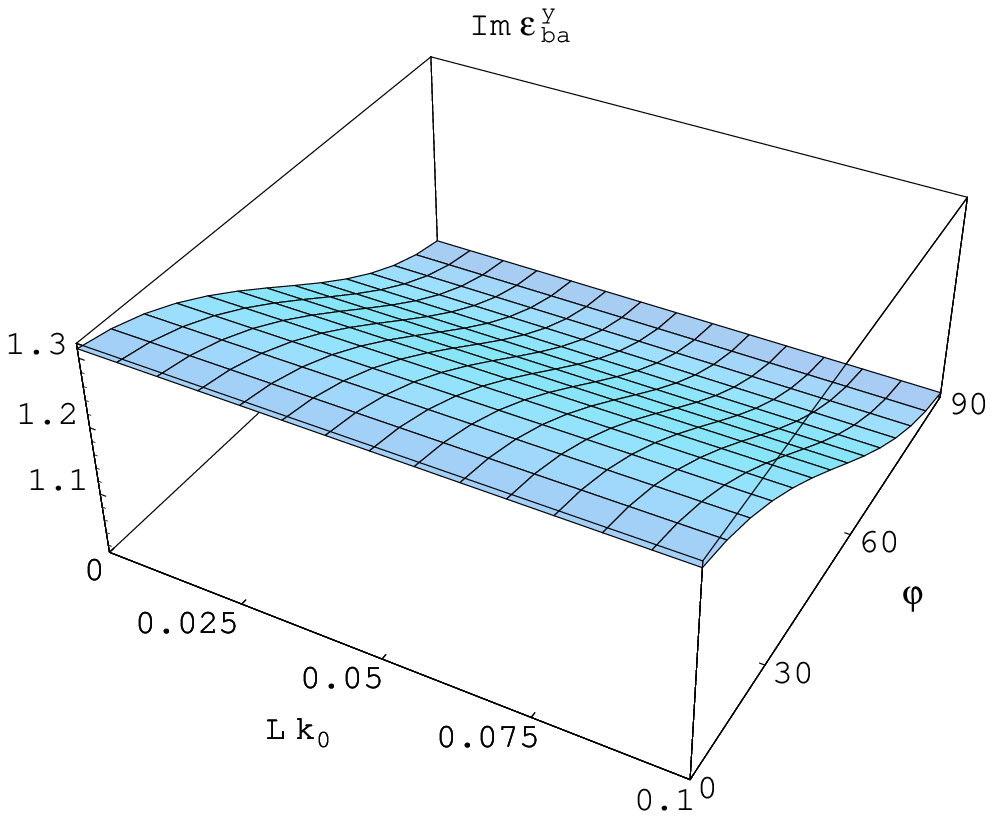,width=2.5in}
\epsfig{file=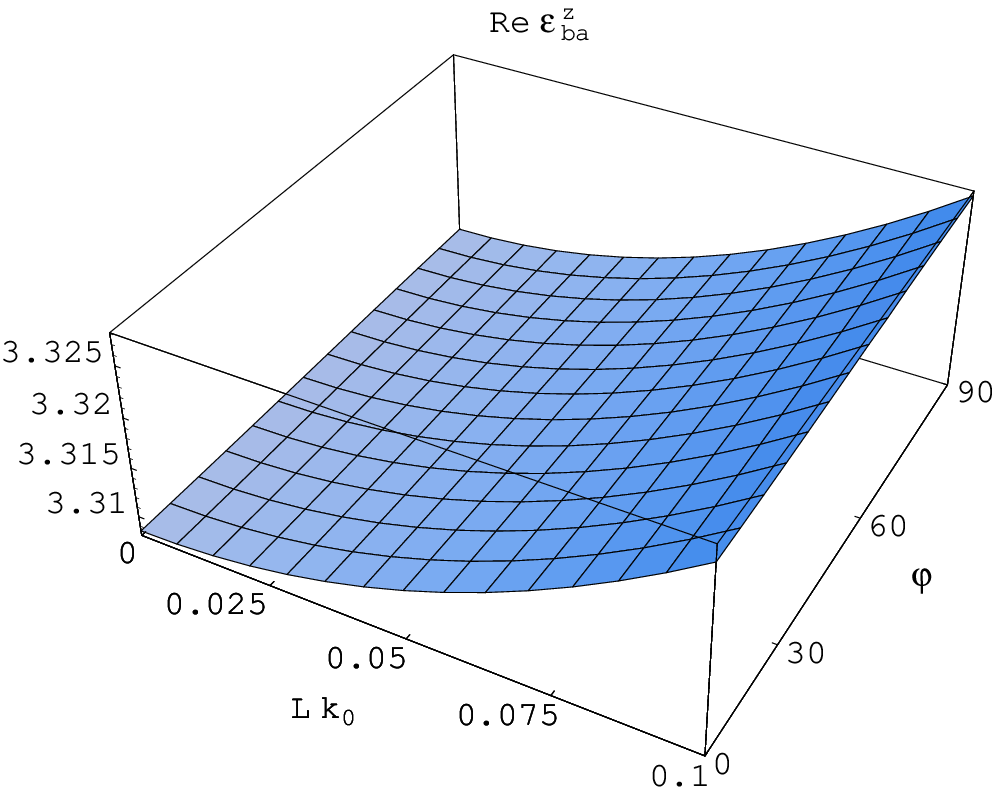,width=2.5in} \hfill
\epsfig{file=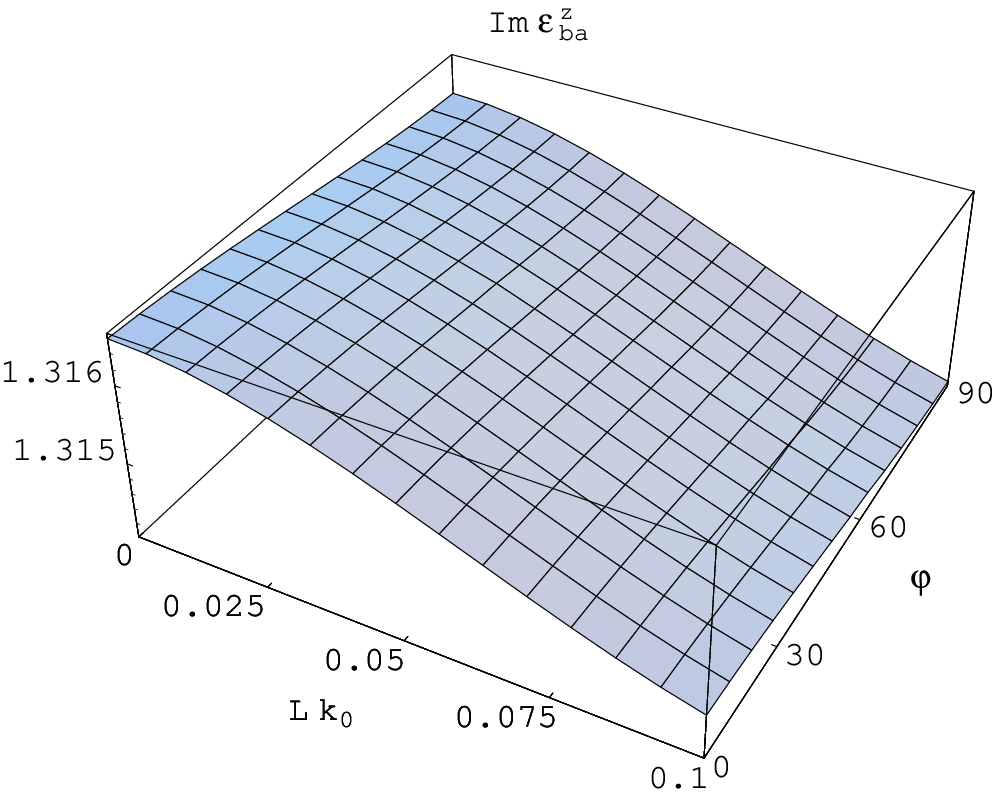,width=2.5in}
\epsfig{file=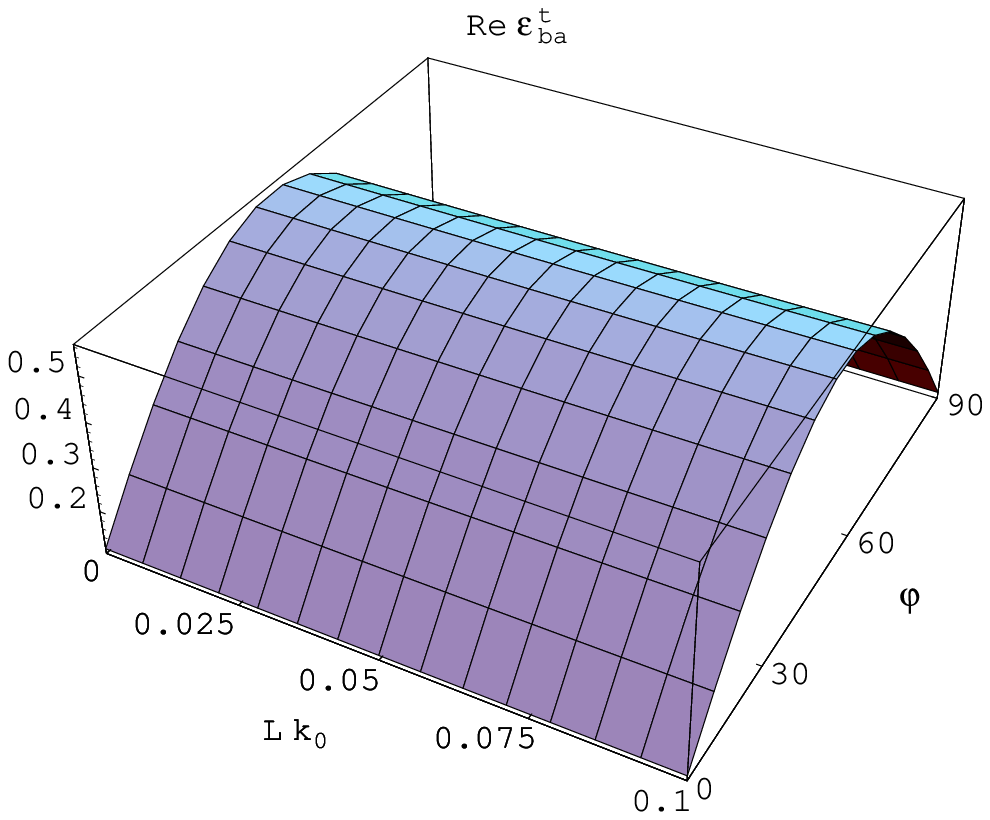,width=2.5in} \hfill
\epsfig{file=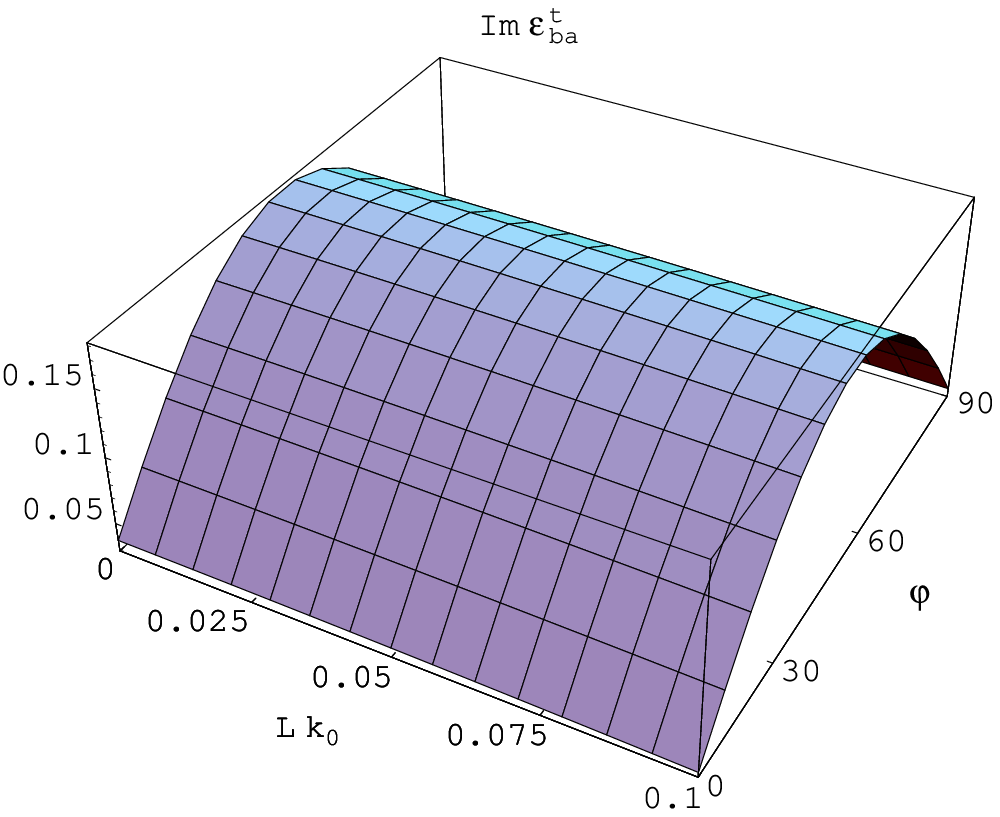,width=2.5in}
 \caption{\label{fig1}
The real (left) and imaginary (right) parts of the components of
 $ \underline{\underline{\hat{ \epsilon}}}_{\,ba}
(0,0,0)\equiv
\underline{\underline{\epsilon}}_{\,ba} $ plotted against the
relative correlation length $L k_{\scriptscriptstyle 0}$ and
orientation angle $\varphi$ (in degrees) of component phase $a$.
The dissipation parameter $\delta = 1$.}
\end{figure}

Consider the four relative permittivity scalars,
$\eps^{x,y,z,t}_{ba}$,  in the unrotated reference frame, i.e., $
\={\hat \eps}_{\,ba} (0,0,0) \equiv
\=\eps_{\,ba} $. The $\eps^{x,y,z,t}_{ba}$ values calculated with
the  dissipation parameter $\delta =1$  are plotted in
  Figure~\ref{fig1}, as functions of
the orientation angle $\varphi$ of component phase $a$ and the
relative correlation length $L \ko$. At $\varphi = 0$ (and also at
$\varphi = \pi$),  the preferred axes of both component phases
are aligned. Accordingly, the HCM is  uniaxial with $\eps^x_{ba} =
\eps^y_{ba}$. For $\varphi \neq 0$ (and also $\varphi \neq \pi$),
the HCM is biaxial.
 As $\varphi \rightarrow \pi/2$, the HCM biaxial structure
becomes orthorhombic since $\eps^t_{ba} \rightarrow 0$. For
intermediate values of $\varphi \in (0,\pi/2)$, the HCM has the
general non--orthorhombic biaxial form \c{MW_biax2}. The
correlation length is found to have only a marginal influence
on $\eps^{x,y,z,t}_{ba}$ for $\delta = 1$.

\begin{figure}[!ht]
\centering \psfull \epsfig{file=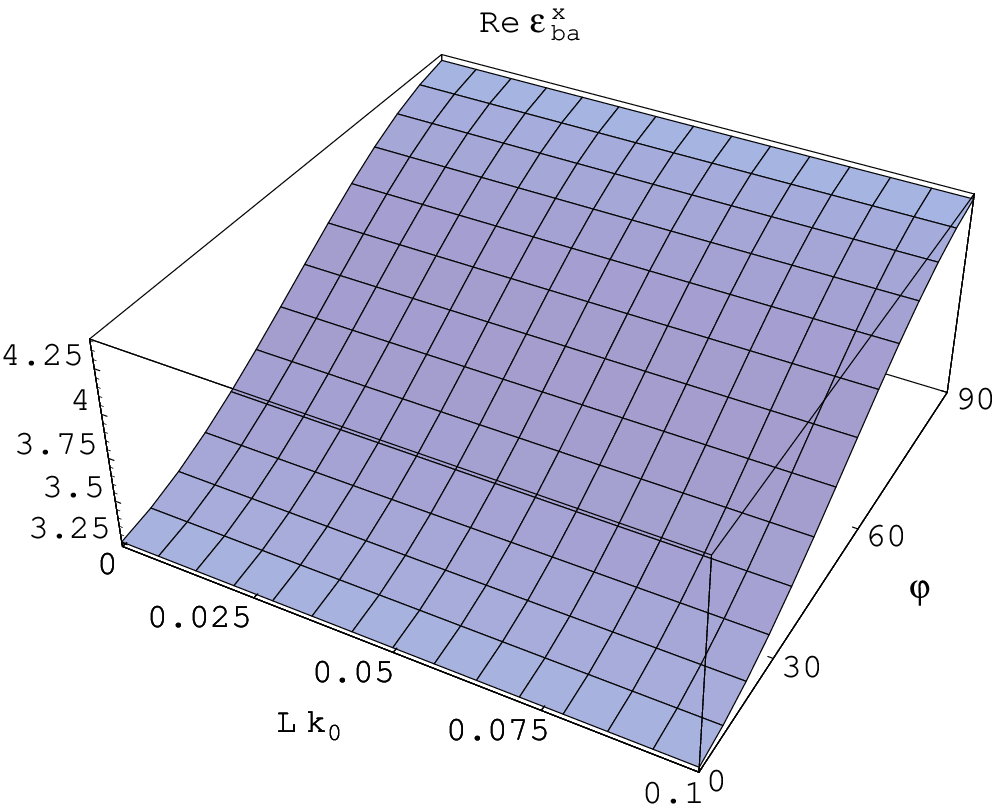,width=2.5in} \hfill
\epsfig{file=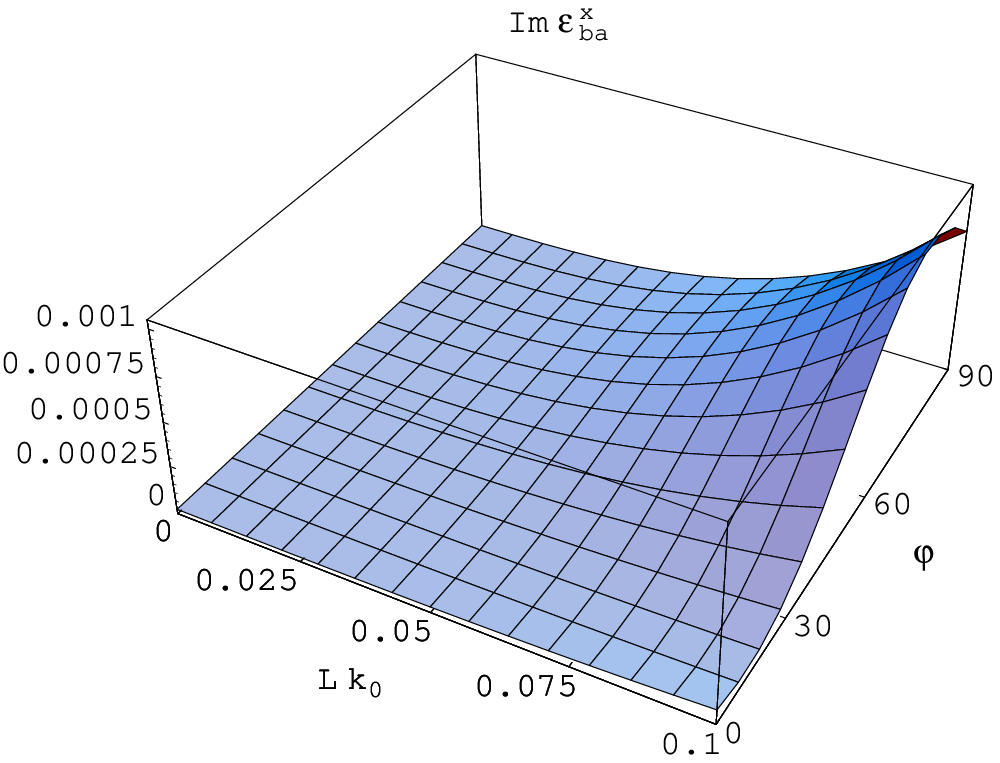,width=2.5in}
\epsfig{file=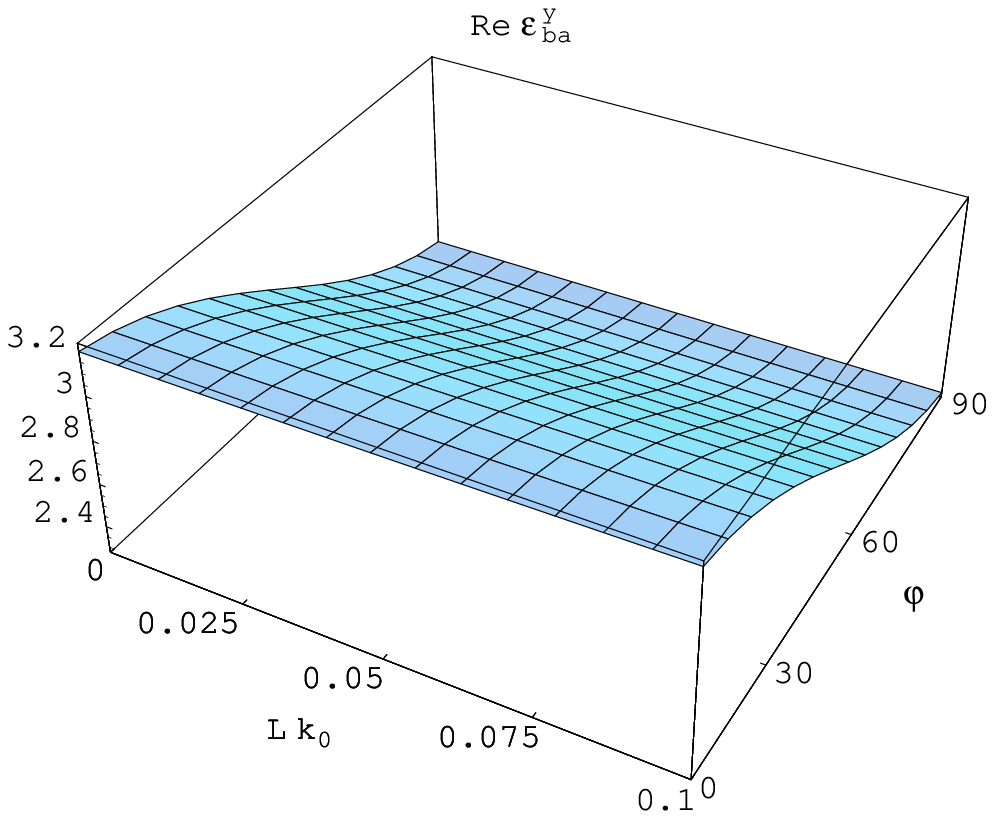,width=2.5in} \hfill
\epsfig{file=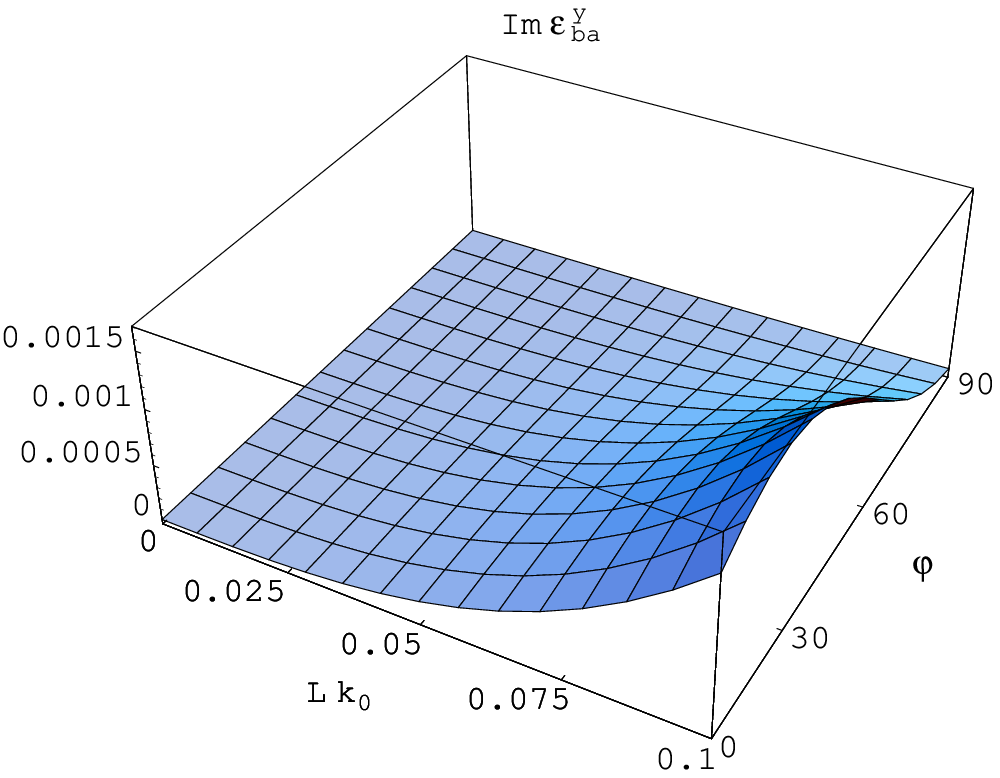,width=2.5in}
\epsfig{file=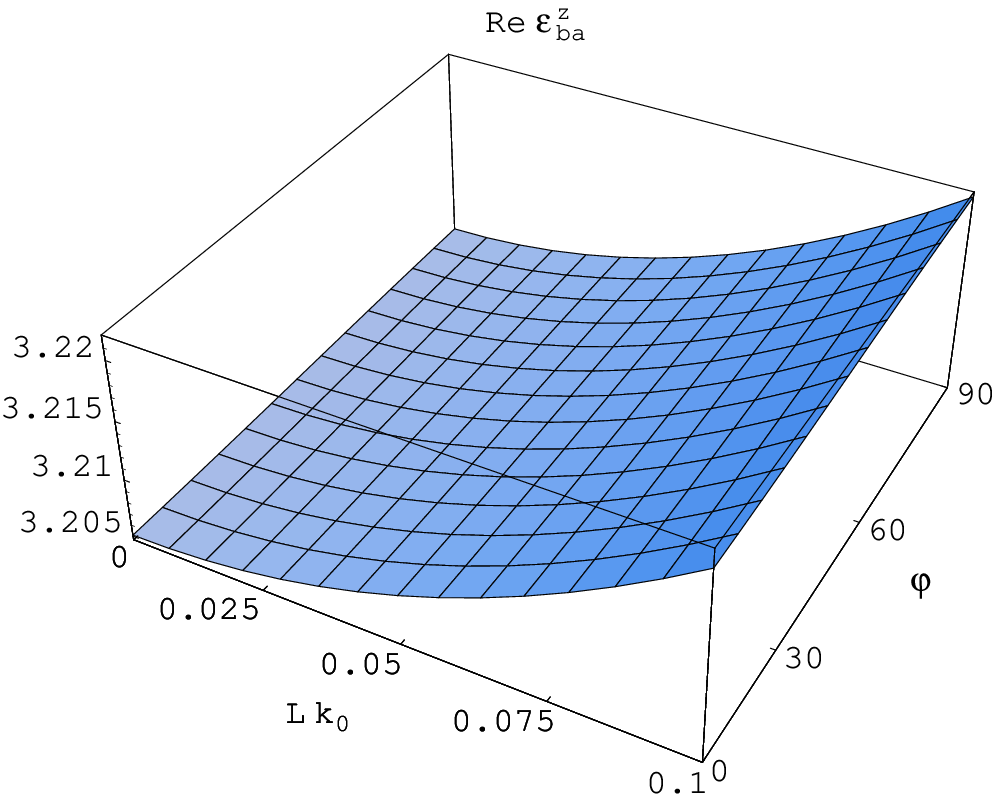,width=2.5in} \hfill
\epsfig{file=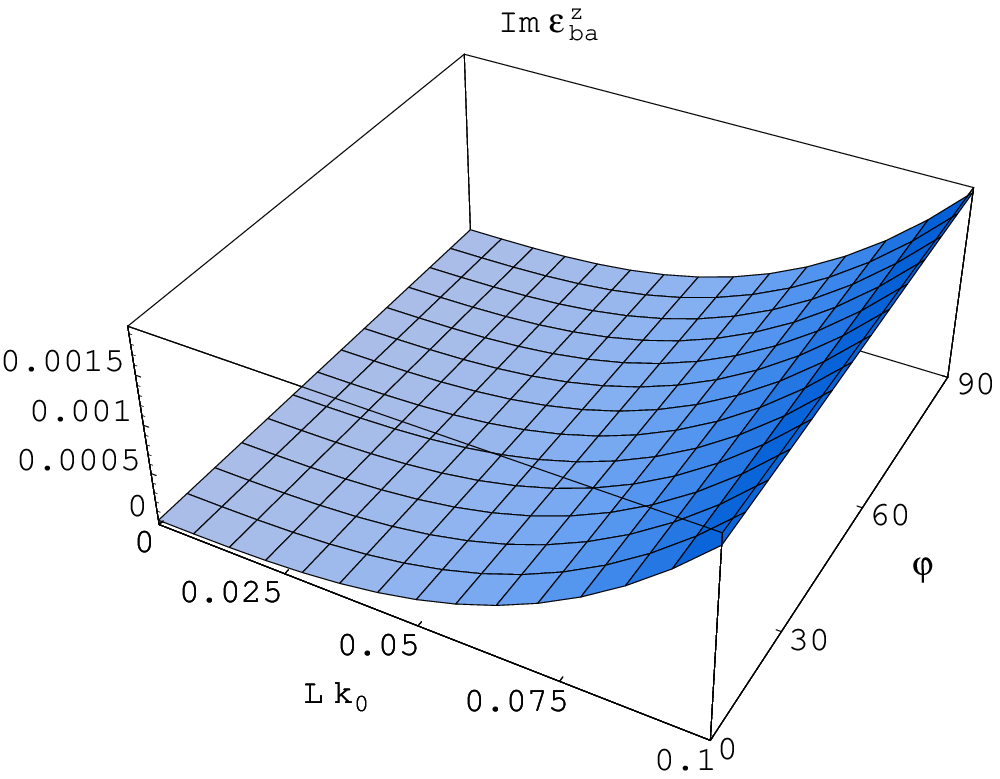,width=2.5in}
\epsfig{file=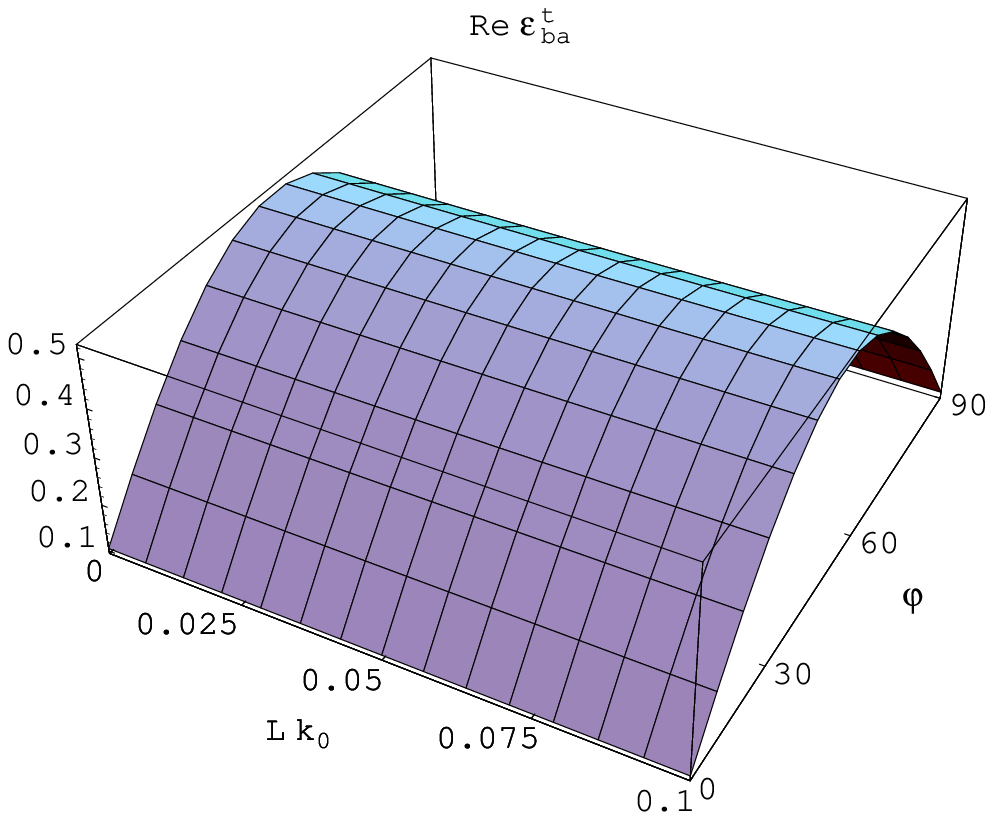,width=2.5in} \hfill
\epsfig{file=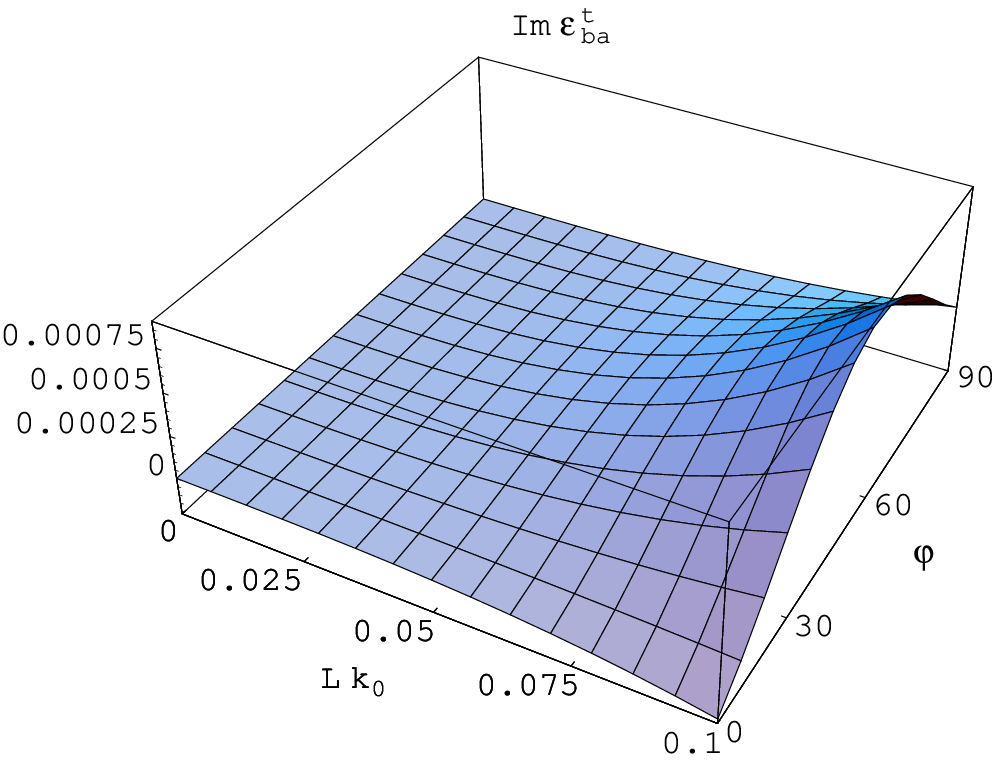,width=2.5in}
 \caption{\label{fig2}
As Figure~\ref{fig1} but with $\delta = 0$. }
\end{figure}

The HCM constitutive parameters  corresponding to those of
Figure~\ref{fig1}, but arising from nondissipative component
phases (i.e., $\delta = 0$), are presented in Figure~\ref{fig2}.
The absence of dissipation in the component phases has little
effect on the real parts of $\eps^{x,y,z,t}_{ba}$. However, the
imaginary parts of $\eps^{x,y,z,t}_{ba}$ are much altered. 
Since the component phases are nondissipative,
 the imaginary parts of $\eps^{x,y,z,t}_{ba}$
are null--valued at zero
correlation length. As the correlation length increases, 
the loss due to the effects of coherent scattering 
becomes greater. Hence, the
magnitudes of the imaginary parts of $\eps^{x,y,z,t}_{ba}$ are observed to
increase in Figure~\ref{fig2} as  $L \ko$ grows.
Furthermore, it is clear from  Figure~\ref{fig2} that
 the rate of increase of these imaginary parts is
sensitively dependent upon the orientation angle $\varphi$ of the
component phase $a$.

\subsection{Zeros of $Y$}

\begin{figure}[!ht] \centering \psfull \epsfig{file=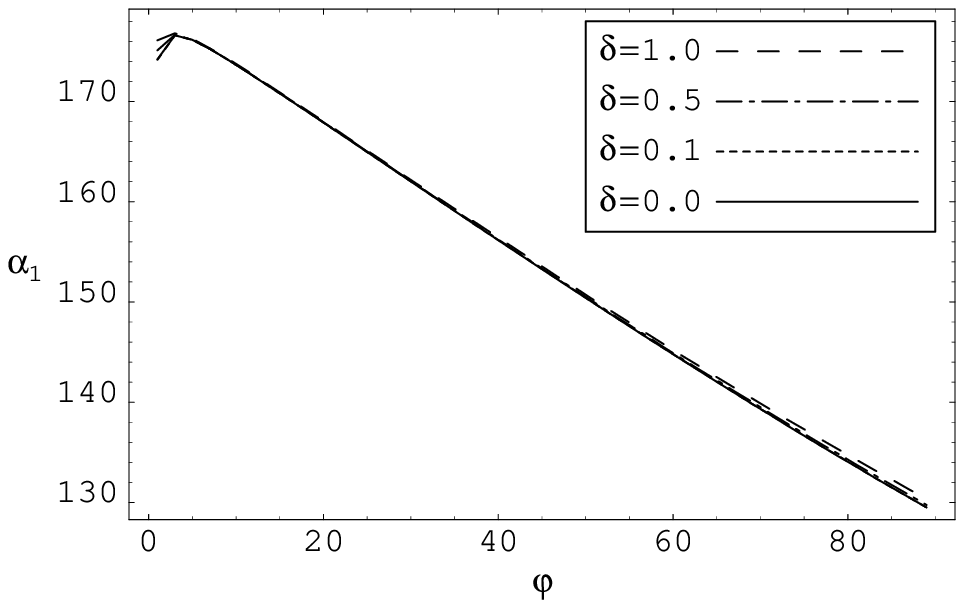,width=3.2in} \hfill
\epsfig{file=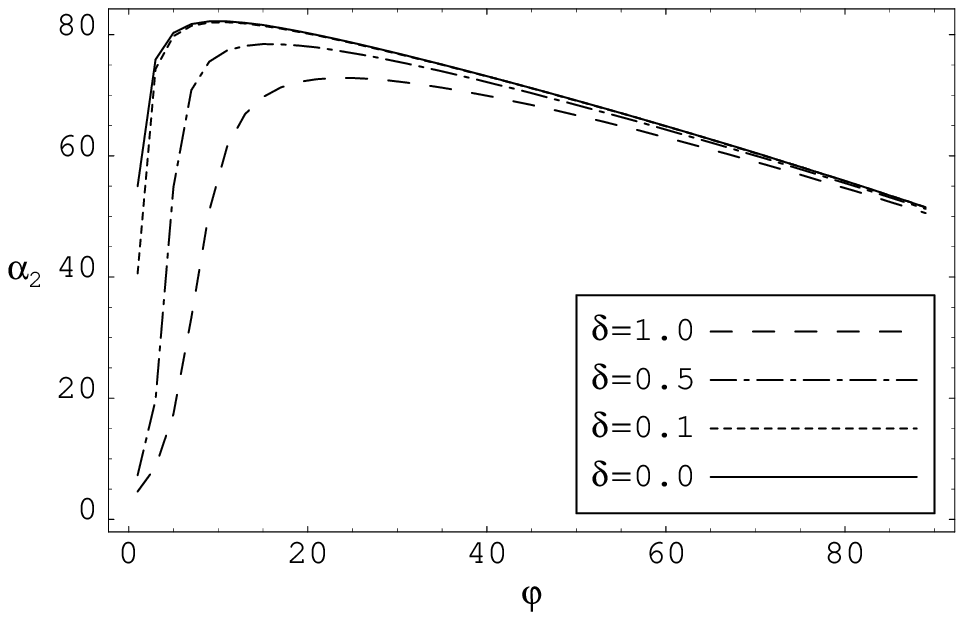,width=3.2in}
\epsfig{file=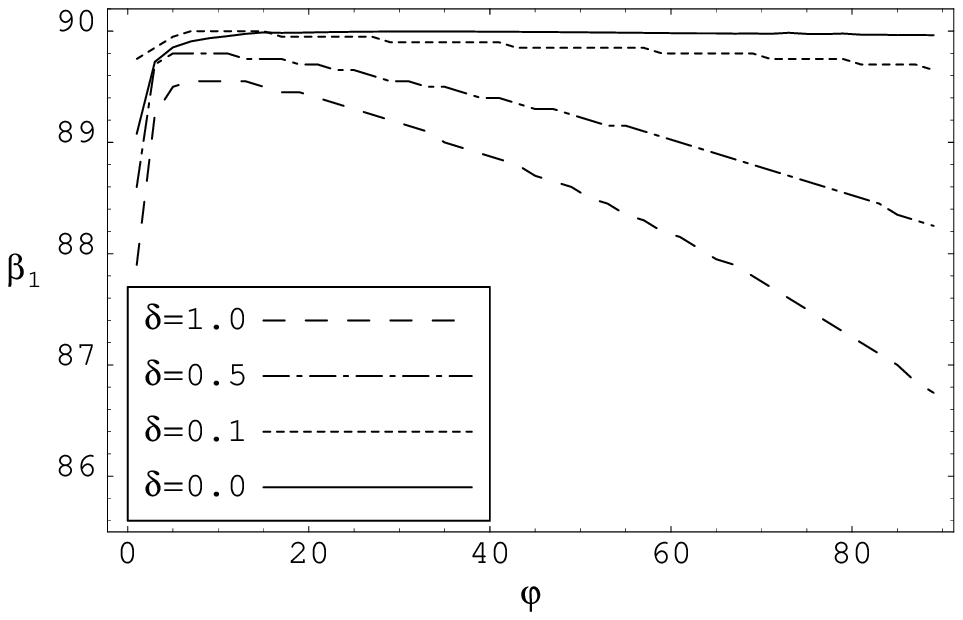,width=3.2in} \hfill
\epsfig{file=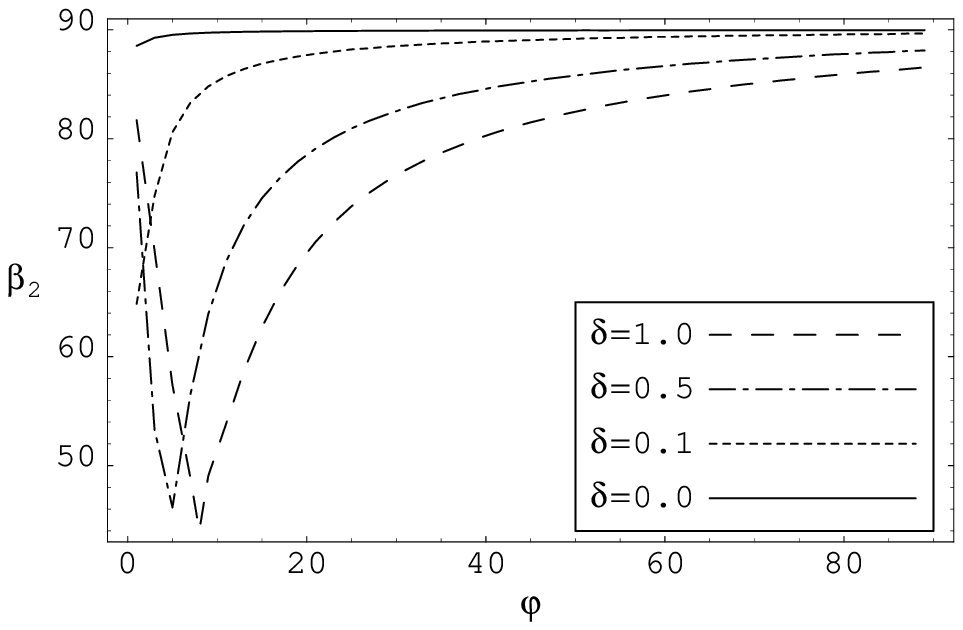,width=3.2in}
 \caption{\label{fig3}
The $\alpha$ and $\beta$ angular coordinates (in degrees) of the
zeros of $Y(\alpha,\beta,0)$ plotted against the orientation angle $\varphi$ (in
degrees) of component phase $a$. The relative correlation length
$L k_{\scriptscriptstyle 0} = 0.1$.}
\end{figure}

The condition $Y = 0$ can be satisfied at two distinct
 HCM
orientations. These orientations  are denoted by
 the angular coordinates $(\alpha_1,\beta_1)$ and $(\alpha_2,
\beta_2)$. With the  normalized correlation length fixed at $L \ko =
0.1$, the  $(\alpha_{1}, \beta_{1})$ and $(\alpha_{2}, \beta_{2})$
angular coordinates are graphed as functions of $\varphi$ in
Figure~\ref{fig3}  for the dissipation parameter values $\delta =
0.0, 0.1, 0.5$ and $1$. In particular, observe that the two
distinct solutions of $Y(\alpha,\beta,0)=0$ exist even when the component material
phases are nondissipative (i.e., $\delta = 0$). The angular
coordinates $(\alpha_1,\beta_1)$ and $(\alpha_2, \beta_2)$ are
clearly sensitive to both $\varphi$ and $\delta$.

\begin{figure}[!ht] \centering \psfull
\epsfig{file=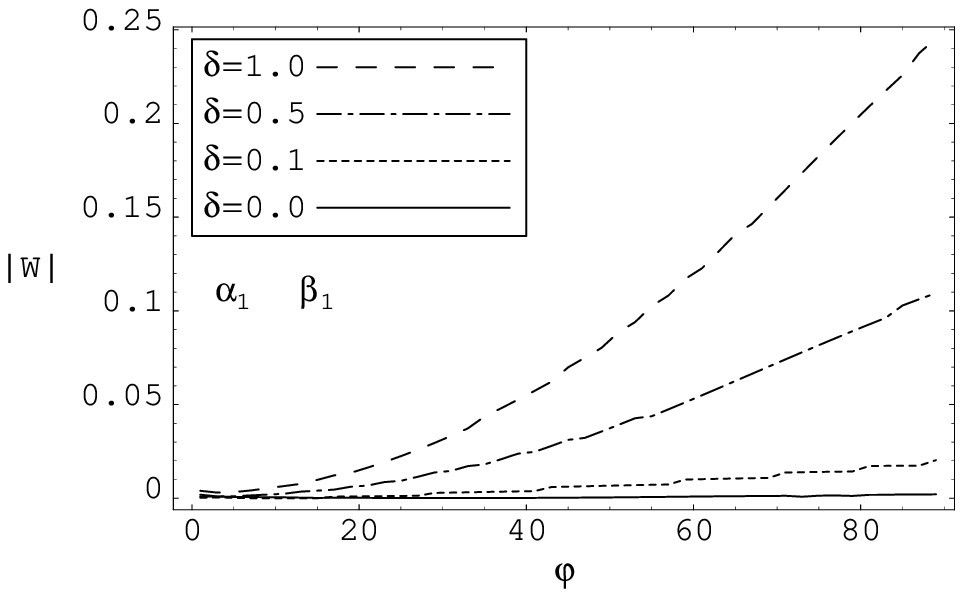,width=3.2in} \hfill
\epsfig{file=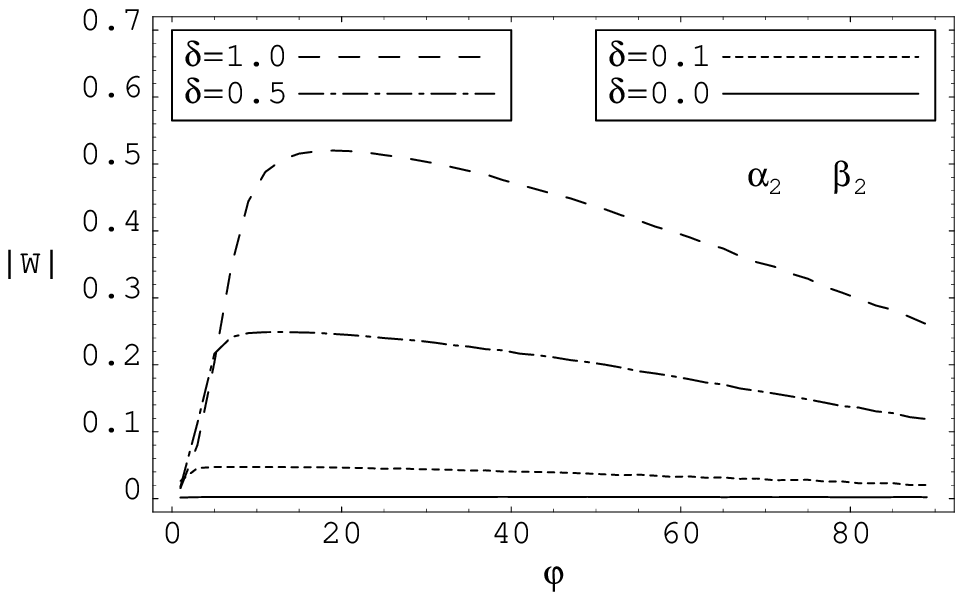,width=3.2in}
 \caption{\label{fig4}
 The values of $|W(\alpha_{1,2},\beta_{1,2},0)|$ corresponding to the $(\alpha_1,\beta_1)$
 and $(\alpha_2,\beta_2)$ angular coordinates
 of Figure~\ref{fig3}, as functions of the orientation angle $\varphi$ (in
degrees) of the component phase $a$.  }
\end{figure}

 Values of $|W|$,  corresponding to the angular coordinates
$(\alpha_1,\beta_1)$ and $(\alpha_2, \beta_2)$ of
Figure~\ref{fig3}, are plotted against $\varphi$ in
Figure~\ref{fig4}. For  $\varphi > 0$, the magnitude $|W| > 0$. In
particular, the inequality $|W| > 0$ holds for $\delta = 0$ (which
is not clearly illustrated in Figure~\ref{fig4} due to limited
resolution). Therefore,  Voigt waves can
propagate along two distinct singular axes in the biaxial HCM, as specified by the
angular coordinates $(\alpha_1,\beta_1)$ and $(\alpha_2,
\beta_2)$, even when the HCM arises from nondissipative component
phases. This conclusion stems solely from the incorporation
of the correlation length in the SPFT, because the Maxwell Garnett
and the Bruggeman formalisms would not predict Voigt wave propagation
when both component phases are nondissipative \c{ML03}.

\begin{figure}[!ht]
\centering \psfull \epsfig{file=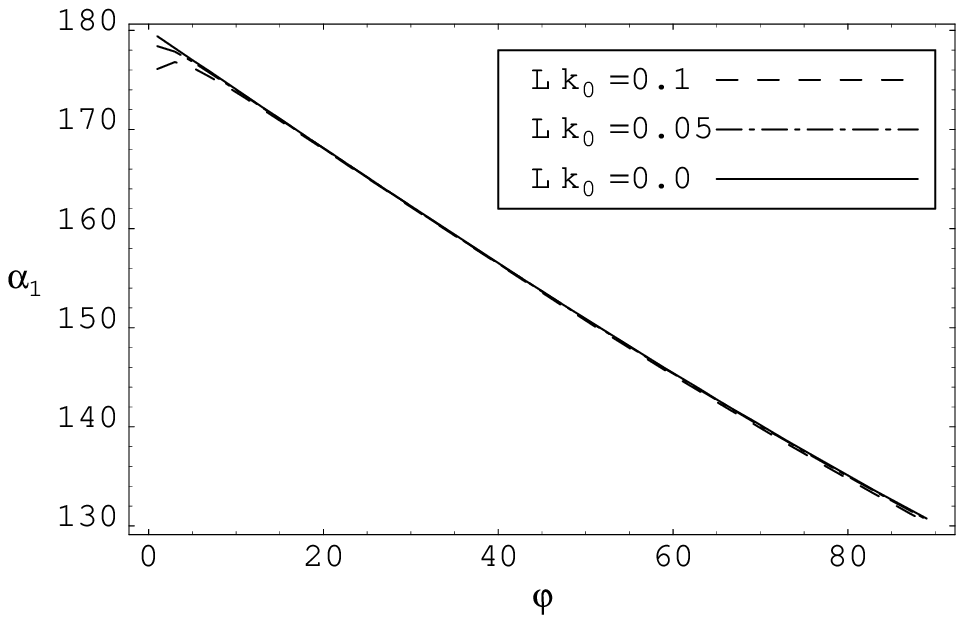,width=3.2in} \hfill
\epsfig{file=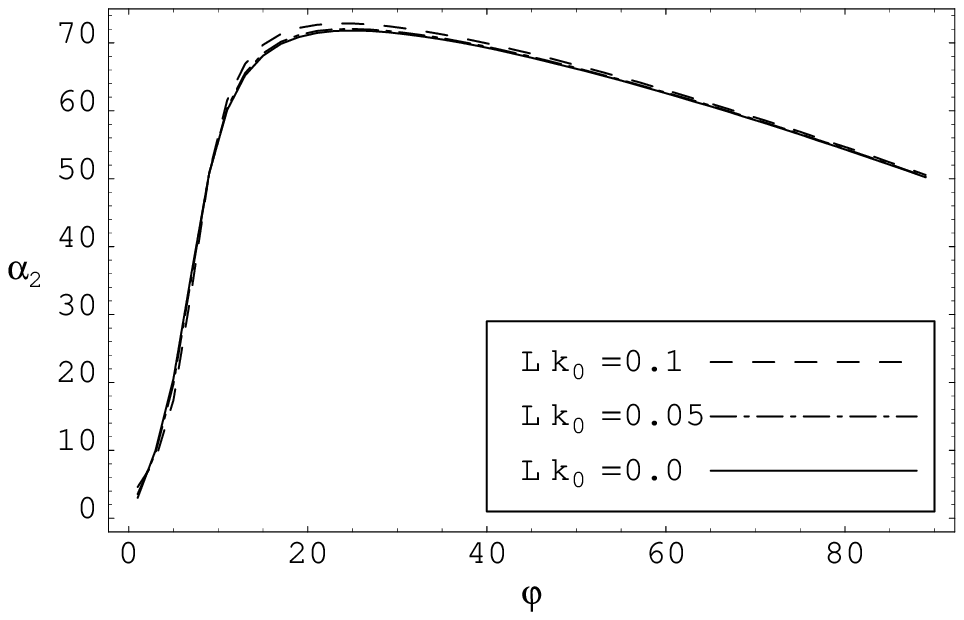,width=3.2in}
\epsfig{file=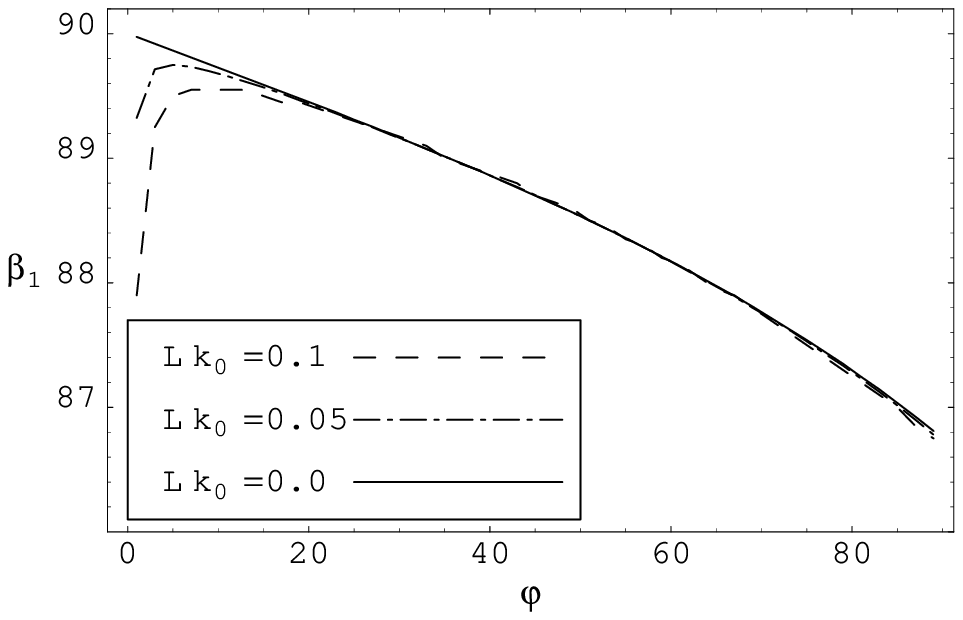,width=3.2in} \hfill
\epsfig{file=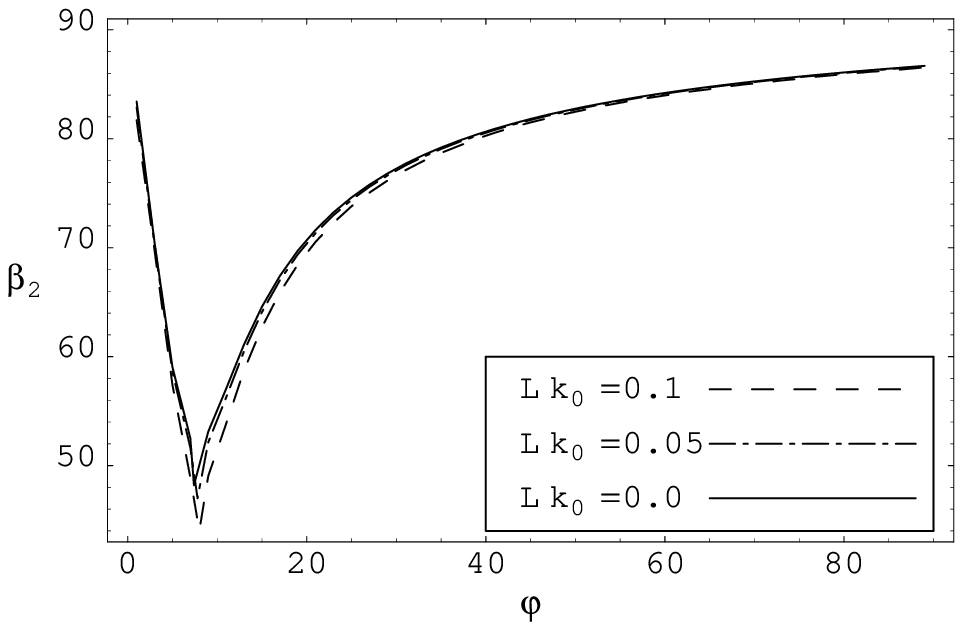,width=3.2in}
 \caption{\label{fig5}
As Figure~\ref{fig3} but with dissipation parameter $\delta = 1$
and relative correlation length $L k_{\scriptscriptstyle 0} = 0,
0.05$ and $0.1$. }
\end{figure}

The two orientations that zero the value of $Y$, as specified
by the angular coordinates $(\alpha_1,\beta_1)$ and $(\alpha_2,
\beta_2)$, are plotted against $\varphi$ in Figure~\ref{fig5} for
the normalized correlation lengths $L \ko = 0, 0.05$ and $0.1$. The
dissipation parameter is fixed at $\delta = 1$. As in
Figure~\ref{fig3}, the two distinct directions described by
$(\alpha_1,\beta_1)$ and $(\alpha_2, \beta_2)$ are sensitively
dependent upon the orientation angle $\varphi$ of the component phase
$a$. Furthermore, the two distinct directions persist in the limit
$L \ko \rightarrow 0$. The influence of   $L \ko$ upon the  angular coordinates $(\alpha_1,\beta_1)$
and $(\alpha_2, \beta_2)$ (as illustrated in Figure~\ref{fig5}) is
relatively minor in comparison with the influence of the
dissipation parameter $\delta$ (as illustrated in
Figure~\ref{fig3}).

\begin{figure}[!ht]
\centering \psfull \epsfig{file=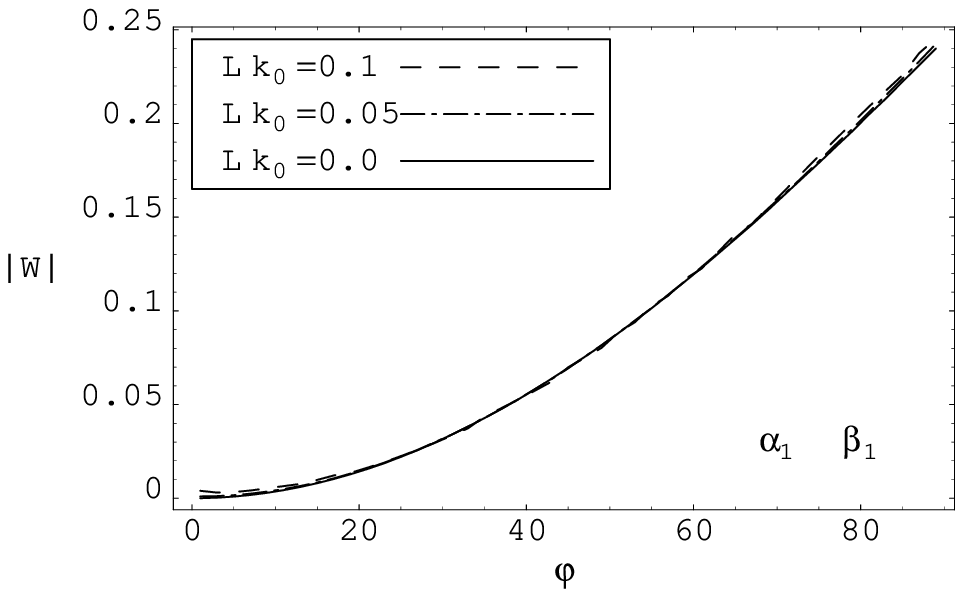,width=3.2in} \hfill
\epsfig{file=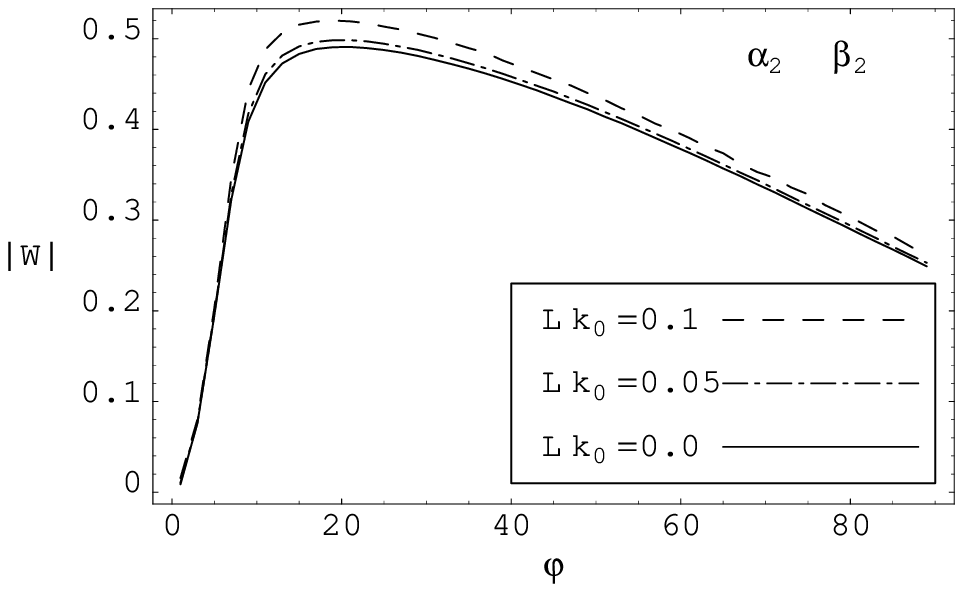,width=3.2in}
 \caption{\label{fig6}
 The values of $|W(\alpha_{1,2},\beta_{1,2},0)|$ corresponding to the  $(\alpha_1,\beta_1)$
 and $(\alpha_2,\beta_2)$ angular coordinates
 of Figure~\ref{fig5}, as functions of the orientation angle $\varphi$ (in
degrees) of the component phase $a$.  }
\end{figure}

For the angular coordinates $(\alpha_1,\beta_1)$ and $(\alpha_2,
\beta_2)$ of Figure~\ref{fig5}, the corresponding values of $|W|$
are presented in Figure~\ref{fig6} as functions of $\varphi$.
Clearly, $|W| > 0$ for $\varphi > 0$ when $L \ko = 0, 0.05$ and
$0.1$. The magnitude of   $L \ko$
has only a minor influence on $|W|$.
 Hence,  the orientations of the
 singular axes, along
which  Voigt waves may  propagate  in the biaxial HCM, are
modulated  only to a minor degree by the correlation length.

\section{Conclusions}

The role of the correlation length in facilitating the propagation
of Voigt waves in HCMs is delineated. Thereby, the importance of
taking higher--order distributional statistics in
homogenization studies  into account is further emphasized. Specifically, we have
demonstrated that

\begin{enumerate}
\item
Voigt waves can propagate in HCMs arising from nondissipative
component phases, provided that a nonzero correlation length is
accommodated, according to the SPFT.

\item The orientations of singular axes in HCMs are sensitively
dependent upon (i) the degree of dissipation exhibited by the
component phases and the  (ii) the orientation of the
preferred axes of the component material phases. By comparison, the
correlation length plays only a secondary role in determining the
singular axis orientations.
\end{enumerate}

\vspace{8mm} \noindent {\bf Acknowledgement:} TGM acknowledges the
financial support of \emph{The Nuffield Foundation}.

\section*{Appendix}

We establish here a surface integral representation of  $\=P$
\r{W_def}, amenable to numerical evaluation. A straightforward
specialization of the evaluation of $\=P$ for bianisotropic HCMs
\c{MLW00} yields the volume integral
\begin{eqnarray}
&& \=P = \frac{f_a f_b}{2 \pi^2 \, i \omega} \int d^3 \#q \;\;
\frac{ (q/\omega)^2 \, \=\alpha + \=\beta}{(q/\omega)^4 \, t_C +
(q/\omega)^2 \, t_B + t_A}\, \le \, \frac{\sin  qL}{q} - L \cos q
L \, \ri\,, \l{W_3d}
\end{eqnarray}
where the scalar quantities $t_A, t_B$ and $t_C$ are given as
\begin{eqnarray}
t_A &=& \muo^3 \eps^z_{Br} \les \,  \eps^x_{Br} \eps^y_{Br} - \le
\eps^t_{Br} \ri^2 \ris, \\
t_B &=&
- \muo^2  \lec \, \eps^z_{Br} \le \eps^x_{Br} + \eps^y_{Br} \ri \cos^2
\theta + \les \Gamma^\phi_{Br} \eps^z_{Br} + \eps^x_{Br} \eps^y_{Br} - \le
\eps^t_{Br} \ri^2 \ris \sin^2 \theta \ric,\\
t_C &=& \muo \le \eps^z_{Br} \cos^2 \theta + \Gamma^\phi_{Br}
\sin^2 \theta \ri.
\end{eqnarray}
The dyadic quantities $\=\alpha$ and $\=\beta$ are specified as
\begin{eqnarray}
\=\alpha &=& \=T_{\,B} - \frac{t_B}{t_C} \, \=T_{\,C},\\
\=\beta &=& \=T_{\,A} - \frac{t_A}{t_C} \, \=T_{\,C},
\end{eqnarray}
with
\begin{eqnarray}
\=T_{\,A} &=& \muo^3 \, \mbox{adj} \le \=\eps_{Br} \ri,\\
\=T_{\,B} &=& - \muo^2 \Big\{ \, \les \eps^z_{Br} + \le \eps^y_{Br} -
\eps^z_{Br} \sin^2 \phi \ri \sin^2 \theta \ris \, \hat{\#x}\,
\hat{\#x}
\nonumber \\ &&
+  \les \eps^z_{Br} + \le \eps^x_{Br} -
\eps^z_{Br} \cos^2 \phi \ri \sin^2 \theta \ris \, \hat{\#y}\,
\hat{\#y}
\nonumber \\ &&
+ \les \le \eps^x_{Br} + \eps^y_{Br} \ri \cos^2 \theta
 + \Gamma^\phi_{Br} \sin^2 \theta \ris \, \hat{\#z}\,
\hat{\#z}
\nonumber \\ &&
 + \les  \le  \eps^z_{Br} \sin \phi \cos \phi - \eps^t_{Br} \ri \sin^2
\theta \ris \, \le
\hat{\#x}\, \hat{\#y}+ \hat{\#y}\, \hat{\#x} \ri
 \, \Big\},\\
\=T_{\,C} &= & \muo \Big[ \sin^2 \theta \cos^2 \phi \, \hat{\#x}\, \hat{\#x} +
 \sin^2 \theta \sin^2 \phi \, \hat{\#y}\, \hat{\#y} \nonumber \\ &&
+ \cos^2 \theta \, \hat{\#z}\, \hat{\#z} +
\sin^2 \theta \cos \phi \sin \phi \, \le \hat{\#x}\, \hat{\#y} +
\hat{\#y}\, \hat{\#x} \ri
\Big].
\end{eqnarray}

Let
\begin{eqnarray}
&& \rho_\pm = \omega^2 \, \frac{ - t_B \pm \sqrt{ t^2_B - 4 t_A t_C}}{2\, t_C}.
\end{eqnarray}
In the long--wavelength regime, i.e.,  $|\,L \sqrt{\rho_\pm}\,|
\ll 1$ \c{MLW00}, the  application of residue calculus to \r{W_3d}
delivers the surface integral
\begin{eqnarray}
 \=P &=&
\frac{L^2 f_a f_b \, \omega}{4 \pi i}
\int^{2 \pi}_{0}  d \phi  \int^{\pi}_0
 d \theta\;\;
\frac{\sin \theta}{3 \, \sqrt{ t^2_B - 4 t_A t_C}} \times \nonumber \\ &&
\Bigg\{ \, \frac{1}{\omega^2} \les \,
 \frac{3}{2}\, \le \, \rho_+ -
\rho_- \, \ri + iL \le \, \rho_+ \sqrt{\rho_+} - \rho_- \sqrt{\rho_-} \, \ri
\ris
  \=\alpha
 +
i L \,\le \, \sqrt{\rho_+} - \sqrt{\rho_-} \, \ri  \=\beta \,
\Bigg\}. \l{surf}
\end{eqnarray}
Standard numerical techniques may be straightforwardly applied to
evaluate \r{surf}.

\newpage

\newpage

\newpage

\newpage

\begin{thebibliography}{99}
\nonumber

\bibitem{Wals}
Walser R M 2003 Metamaterials
 \emph{Introduction to Complex Mediums for Optics and
Electromagnetics} ed WS Weiglhofer and A Lakhtakia (Bellingham,
WA, USA: SPIE Optical Engineering Press) (in press)


\bibitem{ML03}
Mackay TG and Lakhtakia A 2003 Voigt wave propagation in biaxial
composite materials \emph{J. Opt. A: Pure Appl. Opt.} {\bf 5}
91--95

\bibitem{Voigt}
Voigt W 1902 On the behaviour of pleochroitic crystals along
directions in the neighbourhood of an optic axis \emph{Phil. Mag.}
{\bf 4} 90--97

\bibitem{Lakh98}
Lakhtakia A 1998 Anomalous axial propagation in helicoidal
bianisotropic media \emph{Opt. Commun.} {\bf 157} 193--201

\bibitem{Berry}
Berry MV and Dennis MR 2003 The optical singularities of
birefringent dichroic chiral crystals \emph{Proc. R. Soc. Lond. A}
{\bf 459} 1261--1292


\bibitem{Singh}
Singh O N and Lakhtakia A (ed) 2000
 {\em Electromagnetic Fields in Unconventional Materials and
Structures\/} (New York: Wiley)

\bibitem{GL01}
Gerardin J and  Lakhtakia A 2001 Conditions for Voigt wave
propagation in linear, homogeneous, dielectric mediums
\emph{Optik} {\bf 112} 493--495


\bibitem{Khap}
Khapalyuk A P 1962 On the theory of circular optical axes
\emph{Opt. Spectrosc. (USSR)} {\bf 12} 52--54


\bibitem{L96}
Lakhtakia A (ed) 1996 {\em Selected Papers on Linear Optical
Composite Materials\/} (Bellingham, WA, USA: SPIE Optical
Engineering Press)

\bibitem{TK81}
 Tsang L and Kong J A 1981 Scattering of electromagnetic
waves from random media with strong permittivity fluctuations
\emph{Radio Sci.} {\bf 16} 303--320

\bibitem{Genchev}
Genchev ZD 1992 Anisotropic and gyrotropic version of Polder and
van Santen's mixing formula \emph{Waves Random Media} {\bf 2}
99--110

\bibitem{MLW00}
Mackay T G, Lakhtakia A and Weiglhofer W S 2000
Strong--property--fluctuation theory for homogenization of
bianisotropic composites: formulation \emph{Phys. Rev. E} {\bf 62}
6052--6064 Erratum 2001 {\bf 63} 049901(E)


\bibitem{TKN82}
Tsang L, Kong J A and Newton R W 1982 Application of strong
fluctuation random medium theory to scattering of electromagnetic
waves from a half--space of dielectric mixture \emph{IEEE Trans.
Antennas Propagat.} {\bf 30} 292--302


\bibitem{MLW01b}
 Mackay T G,   Lakhtakia A and  Weiglhofer W S 2001
Homogenisation of similarly oriented, metallic, ellipsoidal
inclusions using the bilocally approximated
strong--property--fluctuation theory \emph{Opt. Commun.} {\bf 197}
89--95


\bibitem{M97}
Michel B  1997 A Fourier space approach to the pointwise
singularity of an anisotropic dielectric medium \emph{Int. J.
Appl. Electromagn. Mech.} {\bf 8} 219--227

\bibitem{Frisch}
Frisch U 1970   Wave propagation in random media
\emph{Probabilistic Methods in Applied Mathematics} Vol. 1 ed A T
Bharucha--Reid (London: Academic Press) pp75--198

\bibitem{Arfken}
Arfken G B and Weber H J 1995 \emph{Mathematical Methods for
Physicists} 4th Edition (London: Academic Press)


\bibitem{MW_biax2}
Mackay T G  and Weiglhofer  W S 2000 Homogenization of biaxial
composite materials: dissipative anisotropic properties \emph{J.
Opt. A: Pure Appl. Opt. } {\bf 2} 426--432



\end{thebibliography}
\end{document}